\documentstyle[12pt]{article}\topmargin=-0.1in\oddsidemargin=5mm

\begin{document}\def\p{\phi}\def\P{\Phi}\def\a{\alpha}\def\e{\varepsilon}
\def\be{\begin{equation}}\def\ee{\end{equation}}\def\l{\label}
\def\0{\setcounter{equation}{0}}\def\b{\beta}\def\S{\Sigma}\def\C{\cite}
\def\r{\ref}\def\ba{\begin{eqnarray}}\def\ea{\end{eqnarray}}
\def\n{\nonumber}\def\R{\rho}\def\X{\Xi}\def\x{\xi}\def\la{\lambda}
\def\d{\delta}\def\s{\sigma}\def\f{\frac}\def\D{\Delta}\def\pa{\partial}
\def\Th{\Theta}\def\o{\omega}\def\O{\Omega}\def\th{\theta}\def\ga{\gamma}
\def\Ga{\Gamma}\def\t{\times}\def\h{\hat}\def\rar{\rightarrow}
\def\vp{\varphi}\def\inf{\infty}\def\le{\left}\def\ri{\right}
\def\foot{\footnote}


\title{Some notes about multiplicity distribution at hadron colliders}
\author{E.A.Kuraev, J.Manjavidze\foot{Permanent address: Inst.of
Phys.,  Tbilisi, Georgia}, A.Sissakian\\ JINR, Dubna, Russia}

\maketitle

\begin{abstract}

The idea that the hard processes are dominate at the very high
multiplicity (VHM) final states creation is considered.  For this
purpose quantitative realization of the Pomeron, DIS and large-angle
annihilation (LAA) mechanism combinations are considered in the pQCD
frame. The phase transition (condensation) in the soft pions system
is described as the alternative to above mechanism. It is shown that
in compared to QCD prediction last one predicts enhancement in the
VHM distribution tail.

\end{abstract}

\section\0

The estimations of an expected multiplicities distribution tails at
LHC energies are offered as possible physical program.
Investigation of the multiplicity distributions was popular since
seventies \C{kai}.  The very high multiplicity (VHM) processes as the
attempt to get beyond standard multiperipheral hadron physics was
considered in \C{lesha}.  The hadron theory based on the local QCD
Lagrangians \C{bfkl} and the experimental consequences was given in
the review papers \C{khose}.

We begin with
general analysis. Let $\sigma_n(s)$ be the cross section of $n$
particles creation at the CM energy $\sqrt{s}$.  We introduce the
generating function:  \begin{equation} T(s,z)=\sum
z^n\sigma_n,s=(p_1+p_2)^2>>m^2.  \end{equation} So, the total cross
section and the averaged multiplicity will be:  \begin{equation}
\sigma_{tot}=T(s,1)=\sum\sigma_n,\sigma_{tot}\bar{n}=\sum n\sigma_n=
\frac{d}{dz}T(s,z)_{z=1}.
\end{equation}

At the same time
\begin{equation}
\sigma_n=\int\frac{dz}{2\pi i z^{n+1}}T(s,z)=
\int\frac{dz}{2\pi i }e^{(-(n+1)\ln z+\ln T(s,z))}.
\end{equation}

Applying the steepest descent method we may determine the
asymptotical behavior of $\sigma_n$ at large $n$. It was shown in
paper of T.D.Lee and C.N.Yang \cite{ly} that the singularities
$z_s$ of $T(z,s)$ in the $z$ plane may be located at $|z|\geq1$ only.
We may distinguish following possibilities at $n\to \infty$:\\
1)$z_s=1$: $\s_n>O(e^{-n})$;\\
2)$z_s=\infty$: $\s_n<O(e^{-n})$;\\
3)$z_s=z_c, 1<z_c<\infty$: $\s_n=O(e^{-n})$.\\
The second type belong to the multiperipheral processes kinematics:
created particles form jets moving with different velocities along
the CM incoming particles.

Another information  is included in

\be
\ln T(s,z)=\sum\frac{(z-1)^m}{m!}c_m.
\ee
For instance, if $c_m=0,~m>1$ we have the Poisson
distribution: $\sigma_n=\s_{tot}\frac{(\bar{n})^n}{n!}
\exp{(-\bar{n})}$. If $c_m=\ga_m(c_1)^m$, i.e. $\ga_m$ is the some
restricted function of $m$, than the so called KNO scaling take
place: $\sigma_n\sim \sigma_{tot}f(n/\bar{n})$. One of the mostly
interesting question: is the KNO scaling really takes place?

It was found in seventieth that  the multiperipheral kinematics
dominates inclusive cross sections $f(s,p_c)$.  Moreover, the
created particles spectra do not depend on $s$ at high energies in
the multiperipheral region:
$$
f(s,p_c)=2E_c\frac{d\sigma}{d^3
p_c}=\int\frac{d t_1 d t_2 s_1s_2\phi_1(t_1)\phi_2(t_2)} {(2\pi)^2
s(t_1-m^2)^2(t_2-m^2)^2},~s_1s_2(-p_{c\bot}^2)=st_1t_2.
$$
Here $s_1=(p_a+p_c)^2,s_2=(p_b+p_c)^2,p_c=\alpha_c p_a+\beta_c
p_b+p_{c\bot}$ and $\phi_i(t_i)$ are the impact factors of hadrons.
So the particle $c$ forgot the details of its creation.
It was found experimentally that the ratio
\begin{equation}
\frac{f(\pi^+p\to\pi_- +...)}{\sigma(\pi^+p)}
=\frac{f(K^+p\to\pi_- +...)}{\sigma(K_+ + p)}=
\frac{f(pp\to\pi_- +...)}{\sigma(pp)}
\end{equation}
is universal \C{risk}. This take place due to the two Pomeron
multiperipheral exchange providing the nonvanishing contribution in
the $s$ asymptotics to the cross section. It was implied that the
Pomeron intercept is exactly equal to one. Just this kinematics leads
to the KNO-scaling \C{kno}.

The asymptotics 1) assumes phase transition \C{kac}. The signal of
creation of exotic state of pions say in the isotopic state with
$I_z=0$ (production of anomalous number of $\pi_0$) in the region of
space of pions Compton wavelength order -- so called pion condensate
-- may lead to the observable effect in multiplicity distributions.
One may expect considerable deviation from the regime $O(e^{-n})$
mentioned above.

Let us demonstrate this reason using almost hand-waving
arguments. The effective pion's lagrangian of S.Weinberg \cite{wein}
$L_{eff}\sim(1-\vec{\pi}^2/f_\pi^2)^{-1},f_\pi=140 MeV,$ regards the
current algebra theorems, describe rather satisfactory the soft pions
interaction. Using the functional integral approach we may consider
the $n$ pion's correlator

$$
\int D\pi
f_\pi^{-2n} \pi^{i_1}(x_1)...\pi^{i_{2n}}(x_{2n})
e^{-L_{eff}(\pi)}.
$$
The associated probability of creation of $2n$ pions may estimated,
assuming that the kinetic part of Lagrangian is negligible,
\begin{equation}
p_{2n}=\int_0^1 dx x^n\frac{1}{1-x}e^{[-\frac{1}{1-x}]}
\sim e^{(-2\sqrt{n})},\quad n\to\infty.
\end{equation}
Note, the $\pi^2\approx 1-\sqrt{f_\pi/n}\to 1$ is essential in this
integral. This means that the potential part of Lagrangian is $\sim
\sqrt{n/f_\pi}\to\infty$ and thus the semiclassical approximation is
valid.

\section\0

The Pomeron is treated as a (infinite) set of particles emitted close
to the CM beams direction (within the small angles of order
$\theta_i\sim 2m_h/\sqrt{s}<<1$). We expect that these type
of particles will not be detected by the detectors since
they are move into the beams pipe. The collider experiment detectors
locate at finite angles $\theta_D\sim 1$ and will measure the
products only of particle $c$ decay.

What will happened when instead of one particle a set of particles
with invariant mass square $s_t$ is created at large angles? Then the
cross section will acquire the type suppression factor
$(m^2/s_t)F(\a_s\ln^2(s_t/s_0))$ with the function
\ba
\s_n=\f{\a_s^2}{s_t}NF_n(\a_s^2\ln^2(\f{s}{s_0})),~
N=(\f{s}{s_0})^\D,
\n\\
\D=\a_P-1=\f{12\ln2}{\pi}\a_s\approx0.55,~\a_s=0.2
\ea
Radiative corrections to the intercept was calculated \cite{fl98} in
recent time. The resulting value is $\Delta\approx 0.2$.

The way to obtain detected large multiplicity is to organize DIS-like
experiments, expecting the large-angle scattered hadrons in the
detectors. Large transfers momenta will be decreasing
by ordinary evolution mechanism to the value of order $m_\pi$
and then the Pomeron mechanism of peripheral scattering of the
created hadrons from the pionization region will start.

What the characteristic multiplicities expected from Pomeron
mechanism with the intercept exceeding unity, $\Delta\sim 0.2$? It is
the quantity of order $(s/m_\pi^2)^\Delta\approx 200$ for
$\sqrt{s}=14 TeV$. This rather rough estimation is in agreement
with the phenomenological analysis of A.Kaidalov \cite{kai}, based
on multi-pomeron exchange in the scattering channel.

\section\0

Let now construct the relevant cross sections. It is convenient
to separate them to the classes\\
a) Pomeron regime (P);\\
b) Evolution regime (DIS);\\
c) Double logarithmic regime (DL);\\
d) DIS+P regime;\\
e) P+DL+P regime.\\
The description of every regime may be performed in terms of
effective ladder-type Feynman diagrams (The set of relevant FD
depends on the gauge chosen and include much more number of them).

For the pure Pomeron regime \cite{bfkl} the estimated cross
section have the form: ($y=\f{\alpha_s}{16\pi^2},~\quad m^2\sim
s_0\sim m_\pi^2$)
\begin{eqnarray}
d\s_{2\to 2+n}=\frac{1}{64\pi^2}
\int_{m^2/s}^1\frac{d\beta_n}{\beta_n}\int_{m^2/s}^{\beta_n}
\frac{d\beta_{n-1}}{\beta_{n-1}}\cdots
\n\\\t
\int_{m^2/s}^{\beta_2}\frac{d\beta_1}{\beta_1}
d Z_n \int\frac{d^2q_{n+1}}{(q_{n+1}^2-m^2)^2}(\Gamma_1\Gamma_2)^2,
\\ \nonumber
d Z_n=y^n \Pi_{i=1}^{i=n}\int d^2 q_i\Pi_{i=1}^{i=n}
\frac{(s_i/s_0)^{\alpha(q_i)}\gamma_{i,i+1}^2}{(q_i^2-m^2)^2}.
\end{eqnarray}
Performing $\beta$-integration,
\begin{eqnarray}
\int_{m^2/s}^1\frac{d\beta_n}{\beta_n}\int_{m^2/s}^{\beta_n}
\frac{d\beta_{n-1}}{\beta_{n-1}}...
\int_{m^2/s}^{\beta_2}\frac{d\beta_1}{\beta_1}={L^n}{n!},\quad
L=\ln\frac{s}{m^2}.
\end{eqnarray}

Here $q_i=\alpha_i p_2+\beta_i p_1+q_{i\bot}$ is the 4-momentum of the
virtual gluon joining the emitted particles with 4-momenta $k_i,k_{i+1}$,
$s_i=(k_i+k_{i+1})^2$ is their invariant mass square. One should use
here that
\ba
s_1s_2\cdots s_{n+1}=sE_{1\bot}^2\cdots E_{n\bot}^2,~E_{i\bot}^2=m_i^2+
(\vec{q}_i-\vec{q}_{i-1})^2,
\n\\
m^2<<s_i<<s=2p_1p_2,
\ea
where
$$
\alpha(q_i)=\frac{\alpha_s(q_i^2-m^2)}{2\pi^2}\int\frac{d^2k}{(k^2-m^2)
((q_i-k)^2-m^2)},
$$
is the reggeized gluon trajectory. Here we imply the arrangement on
the rapidities of the emitted gluons
\begin{equation}
\frac{m^2}{s}<<\beta_n<<\beta_{n-1}<...<<\beta_1\sim 1.
\end{equation}
The quantity $\Gamma_{1,2}$ may be associated with the formfactors of
the initial hadrons (simply we replace them by the coupling constants
of $g,~g^2=4\pi\alpha_s$) whereas the quantities $\gamma_{i,i+1}$
associated with the effective vertices of transition of two gluons to
the emitted particle. For the case of emission of scalar particle we
have $\gamma_{i,i+1}=m$. For the case of emission of gluon with
momentum $k_i=q_i-q_{i+1}$ we have
\begin{equation}
\gamma_{i,i+1}=g[-(q_i+q_{i+1})-p_2(\frac{2p_1k_i}{p_1p_2}-
\frac{m^2-q_i^2}{p_2k_i})+p_1(\frac{2p_2k_i}{p_1p_2}-\frac{m^2-q_{i+1}^2}
{p_1k_i})].
\end{equation}

\section\0

For the pure deep inelastic case, when one of the initial hadrons is
scattered at the angle $\theta$ have the energy $E'$ in the cms of
beams whereas the another is scattered at small angle and the large
transfer momentum $Q=4EE'\sin^2(\theta/2)>>m^2,$ is distributed to
the some number of the emitted particles due to evolution mechanism
we have \cite{basic}($\theta$ is small):
\begin{eqnarray}
d\s_n^{DIS}=\frac{4\alpha^2E^{{'}2}}{Q^4 M} dD_n dE'd\cos\th, \n\\
dD_n=(\frac{\alpha_s}{4\pi})^n\int_{m^2}^{Q^2}
\frac{d k_n^2}{k_n^2}\int_{m^2}^{k_n^2}\frac{d
k_{n-1}^2}{k_{n-1}^2}\cdots \int_{m^2}^{k_2}\frac{d
k_1^2}{k_1^2}\int_x^1 d\beta_1\int_{\beta_1}^1 d\beta_2...  \n\\
 \times\int_{\beta_2}^1d \beta_1
P(\frac{\beta_n}{\beta_{n-1}})...P(\beta_1), \quad
P(z)=2\frac{1+z^2}{1-z},
\end{eqnarray}
where the limits of integrals show the intervals of variation and the
integrand is the differential cross section.Again the rapidities
$\beta_i$ are rigorously arranged as well as the transverse momenta
squared.

\section\0

For the large-angles particles production process the differential
cross section (as well as the total one) fall with cms energy
$\sqrt{s}$.We will consider for definiteness the process of
annihilation of electron-positron pair to $n$ photons \cite{gorsh}:
\begin{eqnarray}
d\s_n^{DL}=\frac{2\pi\alpha^2}{s}dF_n, \n\\
dF_n=(\frac{\alpha}{2\pi})^n\int_0^\rho dy_n\int_0^\rho
d x_n\theta(x_n-y_n)\int_0^{y_n}d y_{n-1}
\n\\\t
\int_0^{x_n}d x_{n-1}\theta(x_{n-1}-y_{n-1})...,
x_i=\ln\frac{q_i^2}{m^2}, \quad
y_n=\ln\frac{1}{\beta_i},\quad \rho=\ln\frac{s}{m^2},
\end{eqnarray}
The similar formulae takes place for subprocess of quark-antiquark
annihilation into the $n$ large-angle moving gluons. We note that the
quantities $q_i^2$ may vary up to maximal value,$s$ which corresponds
to the emission at large angles.  The total cross section of
annihilation to any number of photons is:
\begin{equation}
\sigma_{tot}(s)=\frac{2(2\pi\alpha)^{3/2}}{s}\frac{2}{x}I_2(x),
x^2=\frac{2\alpha}{\pi}\rho^2.
\end{equation}
We conclude that the differential cross sections of the $n$ particles
production may be presented as a product of factors
$$\f{dq_i^2}{q_i^2} \f{d\b_i}{\b_i}$$ under various assumptions about
transverse momentum $q_i$ and rapidity $\b_i$.

We will suppose that every emitted particle of mass (virtuality) $M$ will
decay and create the number of secondary particles (pions) with the probability
\begin{equation}
dW_n(M)=d n \frac{c}{\bar{n}} e^{-\frac{cn}{\bar n}},
\bar{n}=\ln\frac{M^2}{m_\pi^2}.
\end{equation}
Construct now the cross sections of combined processes.
When the one of the initial particles $h_1$ is scattered on small but
sufficient enough angle to fit
the detectors and other is scattered almost forward the combination
of DIS and Pomeron regimes take place:
\begin{equation}
d\sigma_{n,m}=d\sigma_n^{DIS} dZ_m,|q_n|^2\sim m^2
\end{equation}
provided that the virtuality of the last step of evolution regime of
order of hadron mass. For the kinematical case of almost forward
scattering of both initial hadrons the situation may be realized with
large angles hadron production from the central region:
\begin{equation}
d\sigma_{n,m,k}=dZ_n d\sigma^{DL}_m d Z_k.
\end{equation}

\section\0

The possibility of exponential fall down mechanisms violation of the
multiplicity $n$ distribution was discussed. It was mentioned that it
take place due to creation of pion condensate.

There is also another possibility \cite{nikit}. The $k\leq
n$-particle state
$$
|k>=\f{1}{\sqrt{k!}}\sum_{perm}\p^+(x_1)\cdots\p^+(x_k)|0>.
$$
If $|x_i-x_j|\geq1/m_\pi$ then the Bose-correlations are absent and
$<k|k>=k!/k!=1$ because of ortho-normalizability of states. But
all correlations become essential if $|x_i-x_j|<1/m_\pi$, then the
states are not orthogonal and, in result, $<k|k>\sim k!^2/k!=k!$.
The probability of such situation rise with rising multiplicity and
the exponential fall off will be changed on the (factorial?) growth.
If such effect will take place some understanding of the Centauro
cosmic event may be reached.

{\bf Acknowledgments}\\

We are grateful to V.G.Kadyshevski for interest to discussed in the
paper questions.

\end{document}